**Title**
- Plasmonic instabilities and terahertz waves amplification in graphene metamaterials.
- Terahertz waves amplification in graphene.

**Authors**

Stephane Boubanga-Tombet,[1*] Deepika Yadav,[1] Wojciech Knap,[2] Vyacheslav V. Popov,[3] Taichii Otsuji[1]

**Affiliations**

1- Research Institute of Electrical Communication, Tohoku University, Sendai 980-8577, Japan.

2- Laboratory Charles Coulomb, University of Montpellier and CNRS, Place Eugene Bataillon, F-34095 Montpellier, France.

3- Kotelnikov Institute of Radio Engineering and Electronics (Saratov Branch), Russian Academy of Sciences, Saratov 410019, Russia.

[*]Corresponding author. Email: stephanealbon@hotmail.com.

**Abstract**

Plasmon oscillations have been intensively studied for more than forty years in conventional two-dimensional electron gas systems in order to find new alternatives to the vacuum devices based on the Smith-Purcell effect in the far-infrared region. However, beside the multiple endeavors, up to date, the plasmon generation in semiconductor heterostructures has been very inefficient. Here we demonstrate that the use of high mobility graphene metamaterials, due to their well-known stronger light-plasmon coupling compared to semiconductor materials can significantly improve the efficiency of far-infrared plasmonic amplifiers and generators. We explore current-driven plasmon dynamics including perfect transparency and light amplification in monolayer graphene structures. Current-induced complete suppression of the graphene absorption is experimentally observed in a broad frequency range followed by a giant amplification (up to ~ 9 % gain) of an incoming terahertz radiation at room temperature. These active plasmonic processes are triggered by relatively low bias voltage in the graphene devices leading to external quantum efficiency of about two orders of magnitude higher than those of the popular optical-to-terahertz conversion devices largely used in far-infrared technologies. Our results combined with the relatively low level of losses and high degree of spatial confinement of plasmons in graphene will open pathways for a wide range of integrated high speed active optoelectronics devices.

**MAIN TEXT**

**Introduction**

Graphene is promised to a wealth of interesting applications and *graphene plasmonics* is emerging as one of the most viable paths for bringing those applications into reality (1,2). Noble metal plasmonics has over the past years allowed significant progress in a number of fields with impressive applications to ultrasensitive detection down to the single-molecule level (3), improved photovoltaics (4), nanoscale photometry (5) as well as cancer therapy (6). Nevertheless, graphene is currently taking off as one of the most vibrant and promising alternatives in the race for new plasmonic materials (7), especially in the mid to far-infrared regions (8–14).

The generation and amplification of electromagnetic waves by plasmonic instabilities in conventional two-dimensional (2D) electron systems (2DESs) have been actively investigated since 1980. The main idea has been to exploit the radiative decay of grating-coupled 2D plasmons for the realization of compact tunable solid-state far-infrared devices (15–27) and thus develop new alternatives to the vacuum devices, such as traveling wave and backward wave tubes based on the Smith-Purcell effect (28), that face severe difficulties to operate beyond the millimeter-wave region (29). However, after about forty years, we are still a long way from the realization of efficient emitters, amplifiers, and generators based on those plasmonic instability-driven mechanisms. The intensity of radiation reported experimentally so far is too small (16,24–26), the plasmon resonances too broad and not tunable enough (27) to be promising for device applications.

The rise of graphene and the well-established stronger light-plasmon coupling in graphene structures compared to conventional 2DESs in semiconductor materials (30) make this work worth to be revisited. Indeed, graphene plasmon based far-infrared devices have recently been investigated (31–34) and significant improvements on gain modulation (32,34), sensitivity (33), and emission (31) have been reported, yielding the foundational pillars for a future robust graphene based plasmonic technology from mid to far-infrared region. Therefore, to address the efficiency limitation of conventional 2D plasmonic grating coupled emitters and amplifiers, we explore a *graphene plasmonic approach*. We investigate dc current driven plasmonic instabilities in high mobility graphene metamaterials that combine the advantage of an efficient tunable absorber, emitter and amplifier at room temperature. Plasmon modes in our devices are excited in monolayer graphene on hexagonal boron nitride (hBN) with a periodic grating gate structure positioned above the graphene sheets as used elsewhere in 2D semiconductors (22,35–42). The grating gate modulates the incoming electromagnetic wave and defines the plasmonic wave vectors. The samples are fabricated as field effect transistors (see Fig. 1A) with structures featuring an interdigitated dual-grating-gate (DGG). The plasmonic cavities are therefore formed below the gates electrodes and designed with symmetric or asymmetric boundaries (41,42) and electron mobilities around 50.000 $cm^2$/Vs at room temperature. We show that in such high mobility graphene/hBN structures, the light-plasmon interaction leads to strong resonance peak in the absorption spectra at 300 K. We observed current-induced plasmon frequency tuning with red shift of the plasmonic resonance and significant reduction of the graphene absorption up to a transparency regime where the extinction completely vanishes. Beyond this complete suppression of the absorption, the dc current induces amplification of the incoming radiation with a blue shift of the resonances. This plasmonic mediated process is remarkably strong with up to ~ 19 % extinction, while in the amplification regime we report up to ~ 9 % amplification coefficients and plasmon resonance tunability over a wide far-infrared frequency range at 300 K. This is the first experimental observation of such plasmon dynamics with clear transitions from absorption to transparency and then amplification regimes, explained here by the occurrence of current-driven plasmonic instabilities and the resulting plasmon-polariton gain in the graphene DGG structures. The external quantum efficiency extracted from our devices is about ~ 0.1 which is several orders of magnitude higher compared to the largely used optical-to-terahertz conversion devices since the work by Auston at al. (43) on picosecond photo-conducting Hertzian dipoles in 1984, and about two orders of magnitude higher compared to the plasmon-improved photo-conducting devices based on conventional semiconductors (44,45).

**Results**

We examined three samples: two asymmetric DGG alignments (A- DGG1 shown in Fig. 1B and A- DGG2 shown in Supplementary Materials, Fig. S1C) and one symmetric DGG alignment (S-DGG, shown in Fig. 1C). Details on samples description are given in the Supplementary Materials. With the applied gate voltages $V_{g1}$ and $V_{g2}$ each device supports the formation of two different plasmonic cavities (types C1 and C2) below the fingers of Gate 1 and Gate 2 in the DGG device structure (see top part of Fig. 1A). Terahertz time-domain spectroscopy (THz-TDS) was employed to measure the changes in the terahertz (THz) pulses transmitted through the graphene plasmonic cavities of type C1 (C2) when sweeping $V_{g1}$ ($V_{g2}$) and keeping the voltage on the other gate electrode constant at the charge neutral point (CNP) $V_{g2} = V_{CNP2}$ ($V_{g1} = V_{CNP1}$). The source-to-drain voltage ($V_d$) dependent measurements were also conducted at a constant $V_g$. The transmission coefficient at a given $V_g$ and $V_d$ is referred to as T while $T_{CNP}$ is the transmission coefficient at $V_g = V_{CNP}$.

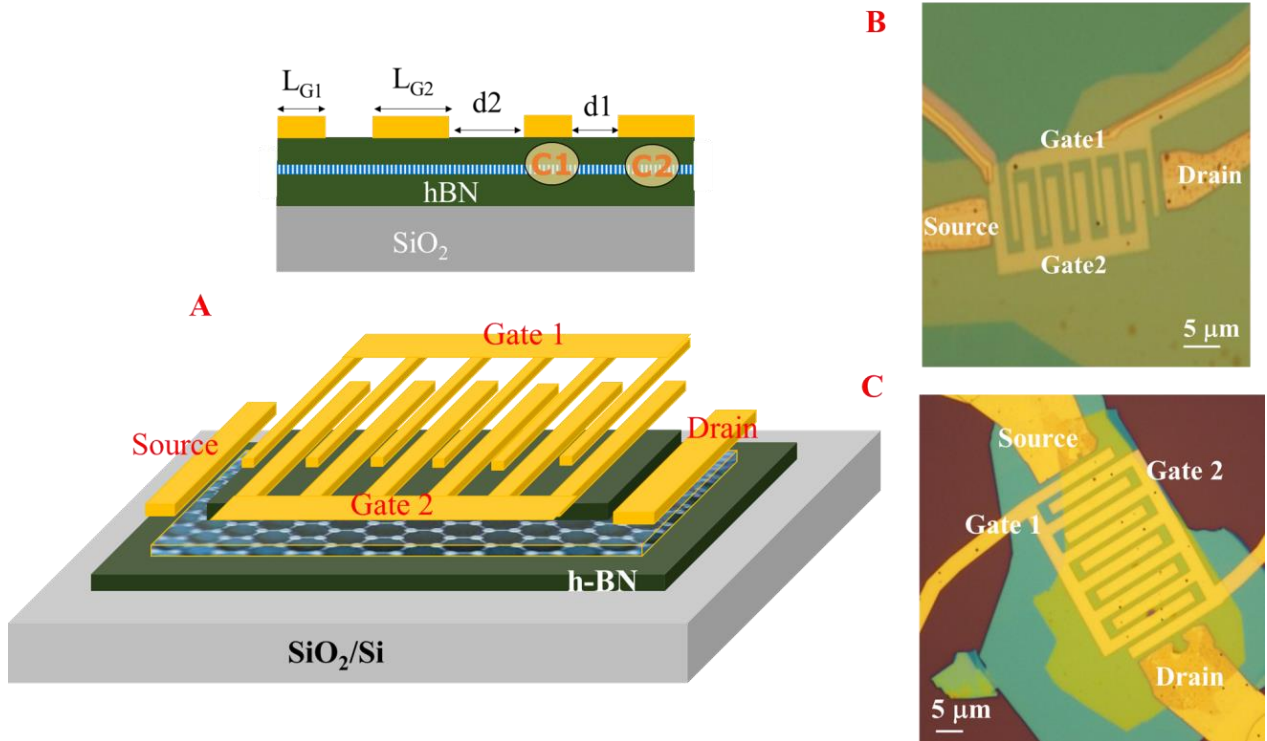

*Figure 1: Schematic view and device images of the active graphene metamaterials. Schematic illustrating of the hBN/graphene/hBN heterostructure including the asymmetric dual-grating-gate metallization (A). Optical image of the asymmetric dual-grating-gate A- DGG1 (B) and the symmetric S-DGG (C) samples.*

Figure 2 depicts the measured extinction spectra (1−T/$T_{CNP}$) in the three devices for $V_d$ = 0 V. The gate length-dependent extinction spectra are shown in Figs. 2, A and B for the same electrical doping of one type of cavities (C1 or C2) $V_g − V_{CNP}$ = 3 V while the other type of cavities (C2 or C1) are in the CNP conditions. The incident THz radiation was polarized parallel (Fig. 2A) or perpendicular (Fig. 2B) to the gates fingers. In the parallel polarization case, the extinction spectra are characterized by the Drude response with a monotonic decrease of the absorption with frequency. In the perpendicular polarization, in contrast, a completely different line shape is seen with a pronounced absorption peak associated to the excitation of the plasmons resonance. We used a damped oscillator and Drude models to describe the line-shape of the extinction spectra (see the Supplementary Materials) shown in dashed lines in Figs. 2, A and B. We observed up to ∼ 19 % absorption in the sample S-DGG ($L_G$ = 2 µm) which is substantially strong for a single atomic layer material, testifying of a strong light to plasmon coupling in our devices. To study the resonance scaling behavior with the carrier density we conducted bias dependent measurements. The gate voltage-dependent extinction spectra are shown in Figs. 2, C and D for the device A-DGG1 with light polarized perpendicular to the gate fingers for different values of $V_{g1}$ while $V_{g2} = V_{CNP2}$ (Fig. 2C) and different values of $V_{g2}$ while $V_{g1} = V_{CNP1}$ (Fig. 2D). It is worth mentioning that throughout all of our experiments, the data from the other devices (A-DGG2 and S-DGG) were quite similar to that shown in Figs. 2, C and D (see the Supplementary Materials). Our gate voltage-dependent data also show a clear blue-shift of absorption peak frequency with increasing $V_g$, as well as an increase of the strength of the plasmon resonance. The scaling behavior of plasmon resonance frequency is shown in Figs. 2, E and F were the resonance peak extracted from the measured extinction are plotted as a function of the plasmon wave vector $q = \pi/L_G$ (Fig. 2E) and gate voltage (Fig. 2F) and fitted to the equation S2 for the dispersion law of gated 2D plasmons in graphene (see the Supplementary Materials). Our data show good agreement between the measured resonance frequencies scaling laws over carrier density and cavity length and the theoretical expectations, allowing us to unambiguously attribute the resonances observed here to 2D plasmons in the graphene/hBN metamaterials. The plasmon absorption coefficients obtained in our graphene structures, with metallic gates-induced plasmonic cavities are rather high, compared to the previously reported results in graphene micro-ribbons (30,46,47), rings and disks (11,48,49).

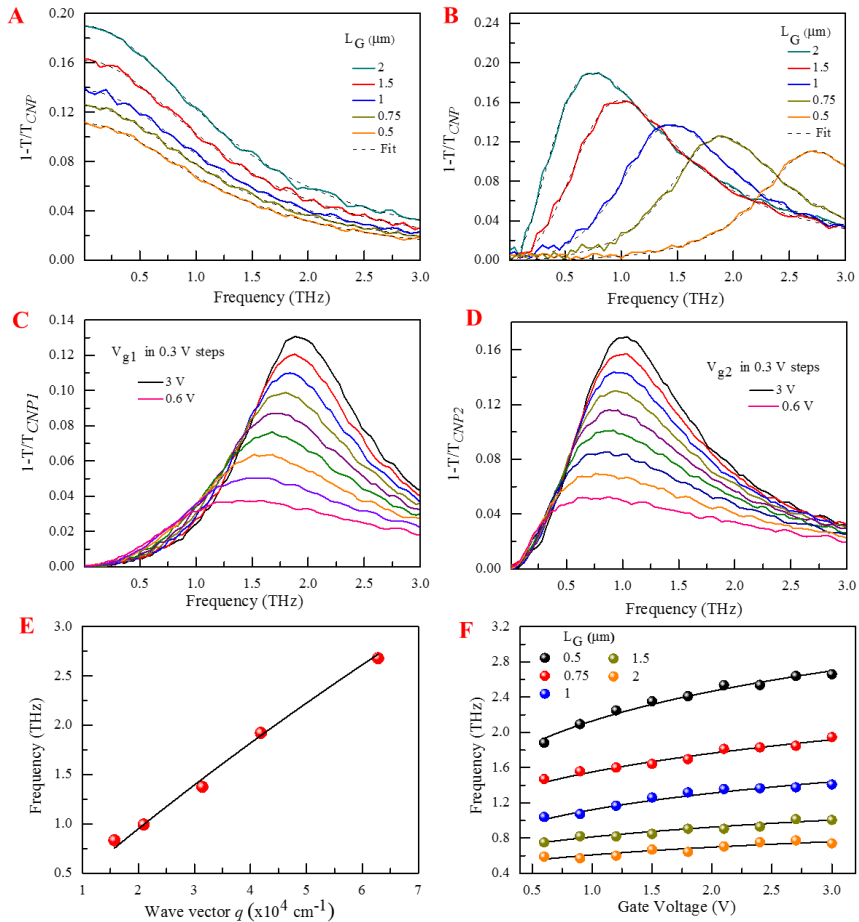

*Figure 2: Gate voltage and length dependent extinction spectra of the graphene structures for $V_d = 0$ V. Plasmonic cavity length ($L_G$) dependent extinction spectra of the three devices with incident light polarized parallel (A) and perpendicular (B) to the gates fingers. The measured line shape is well reproduced by a Drude model fit for parallel polarization (dashed line in (A)) and a damped oscillator model fit for perpendicular polarization (dashed line in (B)). Measured gate voltage-dependent transmission spectra $1-T/T_{CNP}$ of the asymmetric device A-DGG1 with incident light polarized perpendicular to the gates fingers, with biased cavities C1 when sweeping $V_{g1}$ while the voltage on the other gate electrode is kept constant at $V_{g2} = V_{CNP2}$*

*(C) and biased cavities C2 when sweeping $V_{g2}$ while the voltage on the other gate electrode is kept constant at $V_{g1} = V_{CNP1}$ (D). Scaling laws of graphene plasmon resonance frequency in the three devices as a function of wave vector $q = \pi/L_G$ (E) and gate voltage (F). The solid lines in (E) and (F) are fits to data using the model expressed in Eq. S2 (see the Supplementary Materials).*

The most striking features in our measurements arise when we explore the influence of $V_d$ on the devices absorption spectra. Figure 3 depicts the drain voltage- dependent extinction spectra measured in the cavities C1 and C2 of the sample A-DGG1 with the electrical doping at $V_{g1} - V_{CNP1} = 3$ V, $V_{g2} = V_{CNP2}$ (Figs. 3, A and C) and $V_{g2} - V_{CNP2} = 3$ V, $V_{g1} = V_{CNP1}$ (Figs. 3, B and D). As $V_d$ increases the absorption peak clearly shifts to lower frequencies along with a noticeable reduction of plasmon resonance strength. Then the absorption completely vanishes, as seen in the measured extinction spectra being zero at $V_d = 160$ mV (Fig. 3A) and $V_d = 370$ mV (Fig. 3B). The plasmonic devices become perfectly transparent to the incoming THz radiation beam within the entire experimental bandwidth. Here we report the first experimental observation of such a transparency behavior over a relatively wide frequency range (0.1 to 3 THz) in graphene plasmonic structures at 300 K. With increasing $V_d$ beyond this transparency regime, a negative absorption peak appears in the extinction spectra from the lower frequency side with a noticeable blue shift. The negative absorption is an indication of a more intense transmitted pulse as compared to the incoming pulse referred to as gain. This striking influence of $V_d$ can also be visualized in contour plots of the experimental extinction spectra as a function of frequency and $V_d$ shown in Figs. 3, C and D where the lower peak (along the vertical axis) show the device absorption, the upper peak (along the vertical axis) show the gain and the transparency regime between the two peaks with clear current induced transitions between those regions. Throughout all of our experiments, the data from the other asymmetric device (A-DGG2) were quite similar to those shown in Fig. 3 while no amplification was observed in the symmetric sample S-DGG (see the Supplementary Materials).

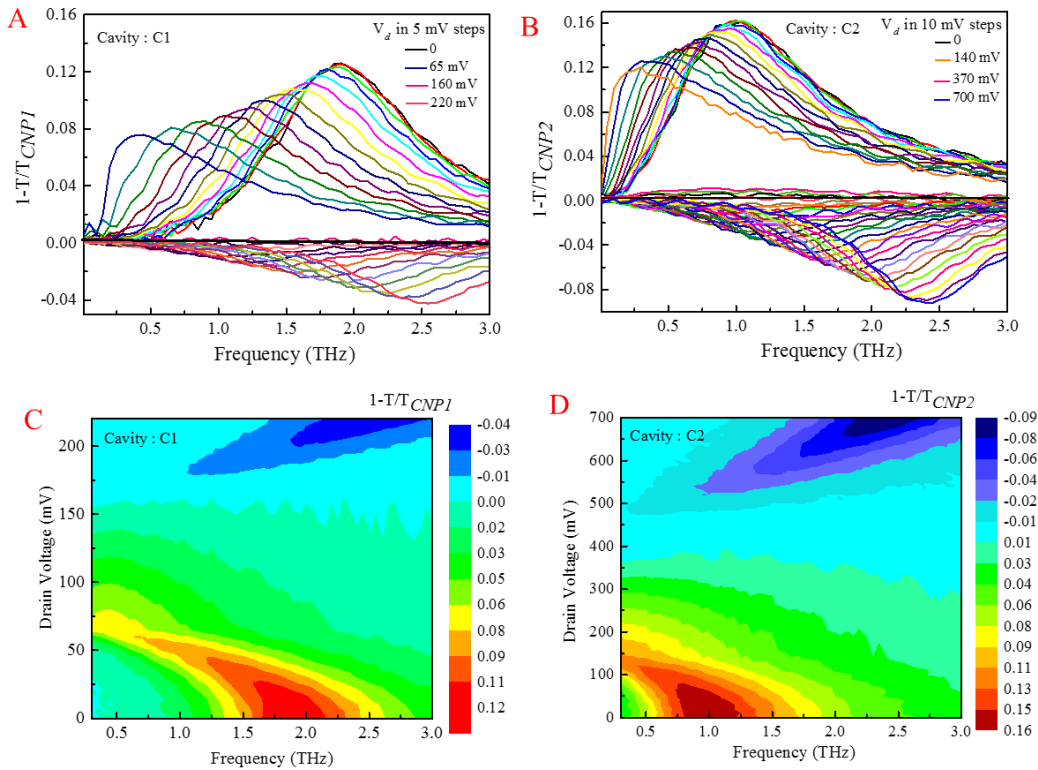

***Figure 3: Drain bias dependent extinction spectra of the graphene structures.*** *Spectra measured*

*in cavities C1 of device A-DGG1 for fixed $V_{g1} - V_{CNP1} = 3$ V and $V_{g2} = V_{CNP2}$ (**A**), and cavities C2 of the same device for fixed $V_{g2} - V_{CNP2} = 3$ V and $V_{g1} = V_{CNP1}$ (**B**) when varying $V_d$. Contour plot of the experimental extinction spectra as a function of frequency and $V_d$ measured in A-DGG1 with the biasing conditions mentioned above for cavities C1 (**C**) and C2 (**D**).*

We now examine the scaling behavior of plasmon resonance frequency with drain bias voltage. The resonance frequencies, $\omega_P(V_d)$ extracted from the drain voltage-dependent extinction spectra for the asymmetric devices are shown in Fig. 4A as a function of the characteristic field intensity $qV_d$ in the plasmonic cavities. By assuming that the carrier drift velocity ($v_0$) is proportional to the mobility and dc electric field in each cavity we estimate that $v_0$ varies from 0 to $\sim 3.85 \times 10^7$ cm/s in our experiments. In the positive absorption region, the plasmon resonance frequencies are described by a scaling behavior of $\omega_P(V_d) = \omega_P(0)[1 - \alpha q V_d]^\beta$ while in the negative absorption region (amplification region) the scaling behavior is $\omega_P(V_d) = \zeta [-1 + \chi q V_d]^\eta$. Our fits (solid lines in Fig.4A) gives $\beta \approx 0.3$ and $\eta \approx 0.5$ while the parameter $\alpha \approx 0.37$ (kV/cm)$^{-1}$ defining the lower cut-off, the same in all cavities, indicates that the onset of the observed transparency regime depends only on the dc electric field in the plasmonic cavities. However, the amplification onset defined by the fitting parameter $\chi \approx 0.098, 0.127, 0.149, 0.187$ (kV/cm)$^{-1}$ for $L_G = 1.5, 1, 0.75, 0.5$ µm varies for different cavities. This cavity length-dependent amplification threshold has been predicted in symmetrical plasmonic cavities based on conventional 2DES (18). In the presence of $V_d$, the source-drain current introduces a collective electron stream flow on which plasma waves are carried at drift velocity $v_0$. Physical phenomena such as Doppler shift of plasmon frequency might therefore occur in the plasmonic system and be considered as possible explanation for the red and blue shift observed experimentally. However, the experimental current-induced tuning range of the resonance frequencies is wider than that expected from a simple classical model. Theoretical approach including the non-trivial carrier hydrodynamics in graphene, the non-uniform density in the channel and asymmetry grating gates remain an open challenge for rigorous quantitative interpretation of the observed drain voltage dependent scaling behavior of plasmon resonance frequencies.

Figure 4B depicts the maximal gain coefficient obtained in the two asymmetric samples as a function of $qV_d$. The applied drain bias voltage varies up to 700 mV with the incoming THz radiation power is of the order of few microwatts. Our data show up to $\sim$ 9 % amplification, which is remarkably high for a single atomic material, testifying of a relatively efficient plasmonic generation process and strong plasmon-radiation coupling. We define the external quantum efficiency as EQE = $P_{out}/P_{in}$ where $P_{out}$ is the device emitted power and $P_{in}$ the injection power. We estimate $P_{out} \approx 0.09$ pW in our graphene based devices while the injection electrical power is $P_{in} \approx 0.7$ pW by considering the upper limit of $V_d = 700$ mV and the protection current limit of $I_p = 1$ µA set to keep the devices safe during experiments. Our devices therefore demonstrate significantly high external quantum efficiency EQE $\approx 0.1$. While relatively inefficient plasmon generation has been reported in semiconductor heterostructures (16,22,24 – 26), the improved efficiency observed in our graphene metamaterials devices is consistent with the well-established stronger light-plasmon coupling in graphene structures compared to conventional 2DES in semiconductor materials (30). Other type of sources operating at 300 K such as the optical-to-THz conversion photoconductive devices are largely used in THz technologies. Recently plasmon based photoconductive devices have shown improved performance in terms of bandwidth and power (44,45) with a maximum emitted power of 250 µW for 80 V bias voltage and 100 mW optical pump power (44) and thus an external quantum efficiency of the order of $\sim 10^{-3}$. Our graphene plasmonic devices show about two orders of magnitude higher external quantum efficiency as compare to those improved optical-to-THz conversion photoconductive devices.

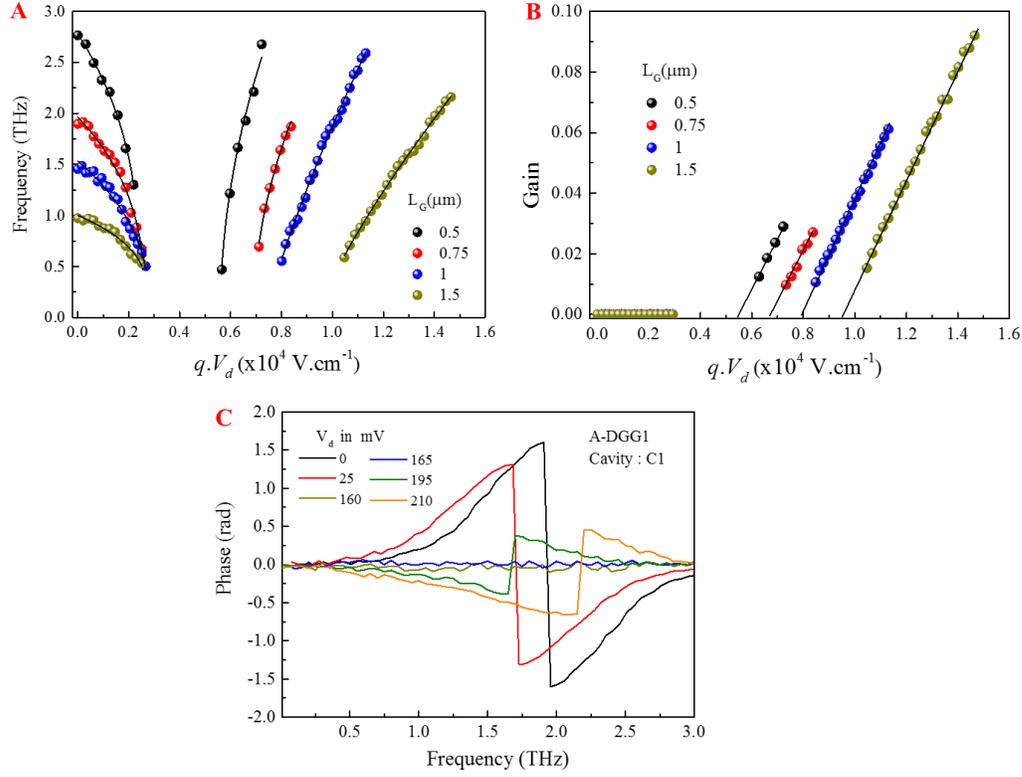

*Figure 4: Drain bias dependent resonance frequencies, gain and phase spectra. Scaling laws of the measured graphene plasmon resonance frequency in the three devices as a function of $qV_d$ and fits to data shown as solid lines (**A**). Maximal gain coefficient (minima of the negative extinction spectra) measured in devices A-DGG1 and A-DGG2 (**B**). Clear threshold like behavior is seen with onset field intensities of ~ 5.4 kV/cm (black dotes), ~6.6 kV/cm (red dotes), ~ 7.8 kV/cm (blue dotes) and ~ 9.4 kV/cm (dark yellow dotes) in the four cavities. Phase change retrieved from the THz electric field in the case of A-DGG1 cavities C1 (**C**). In the region where the negative extinction is observed, the phase shows an inverted behavior compared to that observed in the absorption region.*

The gain is observed beyond threshold field intensities of ~ 5.4 kV/cm (black dotes), ~ 6.6 kV/cm (red dotes), ~ 7.8 kV/cm (blue dotes) and ~ 9.4 kV/cm (dark yellow dotes) in the four cavities. This threshold like behavior indicates that the current-driven plasmonic instability is likely to be the process governing the occurrence of plasmon generation in our graphene/hBN structures. Theoretical investigations of such instabilities have been developed in 2DES based on conventional semiconductor heterostructures (15,17,18,20,21,50,51) and different scenarios are predicted depending on the carrier drift velocity. For values of $v_0$ smaller than the plasma velocity $s$, an instability based on amplified plasmon reflection at the cavity boundaries are predicted in Ref.17, if asymmetrical boundary conditions are realized at different ends of the cavity. When the drift velocity reaches the plasma velocity a different type of instability named dissipative instability (15) can develop along with a choking of electron flow leading to the current saturation (52). Another type of plasmonic instability have been predicted for the carrier drift velocity surpassing the plasmon velocity refereed as plasmonic boom instability occurring with a super-plasmonic boom similar to the well-known supersonic boom (20). Cherenkov-like plasmon instability also may develop in the grating gated 2D electron structure when the electron drift velocity exceeds the plasmon phase velocity (18,21), while in Refs19,53,54 it has been shown that transit time effects in the high field domains of 2DES may drive the plasmonic system unstable. We discuss the possible origins of plasmon instability in our structures by first ruling out the physical mechanisms requiring carrier drift velocity higher than plasmon velocity since the estimated plasmon velocity in our structures $s \sim 3 \times 10^8$ cm/s is higher than the maximum drift velocity $v_0 = 3.85 \times 10^7$ cm/s. Amplified plasmon reflection at the cavity boundaries (17) and transit time effects (19,53,54) are therefore most likely to be the main process governing the observed transparency and gain. Indeed, giant plasmon instability has been predicted in graphene plasmonic structures similar to those reported in the present work, when the amplified reflection and transit time based type of instabilities are superposed with phase synchronization between the neighboring cavities (54).

The common ground within the plasmonic instability picture is the wave amplitude growth with increasing the electron drift velocity (15,17,18,20,21). The plasmon gain also known as instability increment then acts against the plasma wave damping responsible of the losses. The emission is observed above the drift velocity threshold where the gain becomes bigger than the losses. The instability increment in the case of superposition of transit time and amplified reflection types of instabilities is given by (17,19,53,54):

$$\gamma_i = \frac{s}{2L_G}(1-\frac{v_0^2}{s^2})\ln\left|\frac{1+\frac{v_0}{s}}{1-\frac{v_0}{s}}\right| - \frac{u_t}{L_G}\cos(\frac{\pi}{4}\frac{s}{u_t}\frac{L_t}{L_G})J_0(\frac{\pi}{4}\frac{s}{u_t}\frac{L_t}{L_G}) \qquad (1)$$

where $J_0(x)$ is the zeroth order Bessel function, $L_t/u_t$ is the electron transit time from one plasmonic cavity to the other one, $u_t$ and $L_t$ are the drift velocity in the transit regions and the length of those regions, respectively. The decrement is given by $\gamma_d = \Gamma_P/2$, where $\Gamma_P$ is the plasmon resonance line-width, which is ~ 0.6 THz around the lower cut-off defined by the occurrence of the transparency regime. By assuming $u_t \sim$ Fermi velocity ($V_F$) in the depleted regions we obtain $\gamma_i \approx 0.87$ THz. This increment is bigger than the decrement which supports the idea of the two types of instabilities being the governing mechanisms of the observed amplification. While asymmetric boundaries condition is one of the important requirements for the occurrence of the amplified reflection types of instabilities, an additional indication supporting the hypothesis of this type of instability being the governing mechanisms here is the fact that no amplification was observed in our symmetric devices (S-DGG) within the experimental conditions (see the Supplementary Materials. Furthermore, higher plasmon Fourier harmonics with phase velocity smaller than the drift velocity may also play an important role here causing the Cherenkov-like plasmon instability. Indeed, in strongly modulated 2DES the higher Fourier harmonics can be effectively excited even in the lowest frequency plasmon mode (55).

Figure 4C depicts the drain voltage-dependent phase change retrieved from the THz electric field in the case of cavities C1 (see Fig. 3A). In the absorption region ($V_d = 0$ and 25 mV) the phase spectra display the usual behavior for resonant absorption. In the region where negative extinction spectra were obtained ($V_d = 195$ and 210 mV), in contrast, a transition can be seen in the phase spectra which now displays the positive part on the high frequency side of the resonance, inverted compared to the usual behavior for resonant absorption. This transition is an unambiguous evidence for gain and additionally supports our interpretation of light amplification by current driven instability in the high mobility graphene/hBN structures.

**Discussion**

In summary, we have introduced an efficient method to exploit plasmonic instabilities in 2DESs for light generation, and amplification. This approach is based on asymmetric dual grating gates high mobility graphene metamaterial but can potentially be applied in other emerging 2D materials with low level of losses and high degree of spatial confinement of plasmons. This work answers the problems stated about forty years ago about the possibility of using the radiative decay of grating-coupled 2D plasmons for the creation of tunable compact solid-state based far-infrared emitters and amplifiers. The plasmonic instabilities in 2DESs is a reliable method for far-infrared light generation in contrast to optical down-conversion to terahertz frequencies based on nonlinear optical effects and the optical-to-terahertz conversion through photoconduction, the latter present a very low conversion efficiency and the former also is inherently inefficient due to the optical and terahertz phase mismatch limiting the efficient field interaction length. Moreover, while other type of emitters such as quantum-cascade lasers operates exclusively at cryogenic temperatures, all results presented in this work are obtained at room temperatures. The graphene plasmonic solution presented here offer new ways for designing efficient devices for future robust far-infrared plasmonic technology and establish new important challenges for theoretical physics still requiring a physical model allowing full quantitative description of current − driven plasma phenomena in graphene and other 2D systems with Dirac like energy band structure.

**Materials and Methods**

**THz-TDS system**. For the generation of broadband THz waves, a mode locked erbium-doped femtosecond fiber laser with a pulse repetition rate of 80 MHz was used. The pulse width is approximately 80 fs and the central wavelength is 1550 nm. The pulsed laser beam was focused onto a biased, low-temperature-grown InGaAs/InAlAs THz emitter (TERA15-TX, Menlosystems). The antenna had a gap spacing of 100-µm and a bias of 15 V was applied with an average optical power of 20 mW. The free space emitted THz wave was collected and refocused onto the sample (and subsequently the THz receiver for detection) using off-axis parabolic mirrors. A small 250-µm diameter hole at the end of a tapered aperture of a conical shape was placed close to the sample. The propagating THz radiation was detected using a second low-temperature-grown InGaAs/InAlAs THz emitter (TERA15-RX, Menlosystems) made of 25-µm dipole and 10-µm gap excited with an average optical power of 17 mW (at 1550 nm). The THz-TDS system has a usable bandwidth of $0.1-3$ THz and a signal-to-noise ratio above $10^4:1$.

**Sample fabrication**. Graphene and hBN layers were mechanically exfoliated and transferred onto a $SiO_2$/Si wafer (the thickness of $SiO_2$ layer is 200 nm). A hBN/monolayer-graphene/hBN herostructure was realized with the bottom hBN thickness of about 50 nm. Both photolithography and electron-beam lithography were used to define the source, drain electrodes and grating gate patterns. Gold (60 nm) was evaporated to form the electrodes. The gate dielectric consists of a $\sim$ 40-nm-thick hBN layer. The active region of the devices is about 40 µm x 10µm and comprise of 6 periods of wide and narrower gate fingers on top of the hBN gate dielectric. The number of layers of graphene and hBN was determined from color contrast and verified by Raman spectroscopy.

**Supplementary Materials**

section S1. Samples description.

section S2. Supplementary gate voltage results.

section S3. Supplementary drain voltage dependent results.

fig. S1. Device images and the gate dependent resistance of active graphene metamaterials.

fig. S2. Supplementary data, gate voltage dependent extinction spectra of the graphene structures.

fig. S3. Supplementary data, drain bias dependent extinction spectra of the graphene structures.

**Acknowledgments**

**General**: We thank Dr. Dmitry Svintsov, Dr. Akira Satou, Pr. Michel Dyakonov, Pr. Victor Ryzhii and Pr. Michael S. Shur for helpful discussions.
**Funding:** This work was supported by JSPS KAK- ENHI 16H06361 and 16K14243, Japan; VVP acknowledges the support from the Russian Academy of Sciences (Program of Basic Research of the RAS Presidium #1)
**Author contributions:** S. B T. conceived the experiments. S.BT. designed and supervised the fabrication of devices, assisted by T.O. and D.Y. S.BT, and D.Y performed the measurements. S.BT, D.Y, W. K and T.O analyzed the measured data, while V.V.P. contributed to the theoretical explana- tion. All authors discussed the results and contributed to writing the manuscript.
**Competing interests:** The authors declare that they have no competing interests.

**Data and materials availability:** All data needed to evaluate the conclusions in the paper are present in the paper and/or the supplementary information. Additional data related to this paper may be requested from the authors.

**Figures**

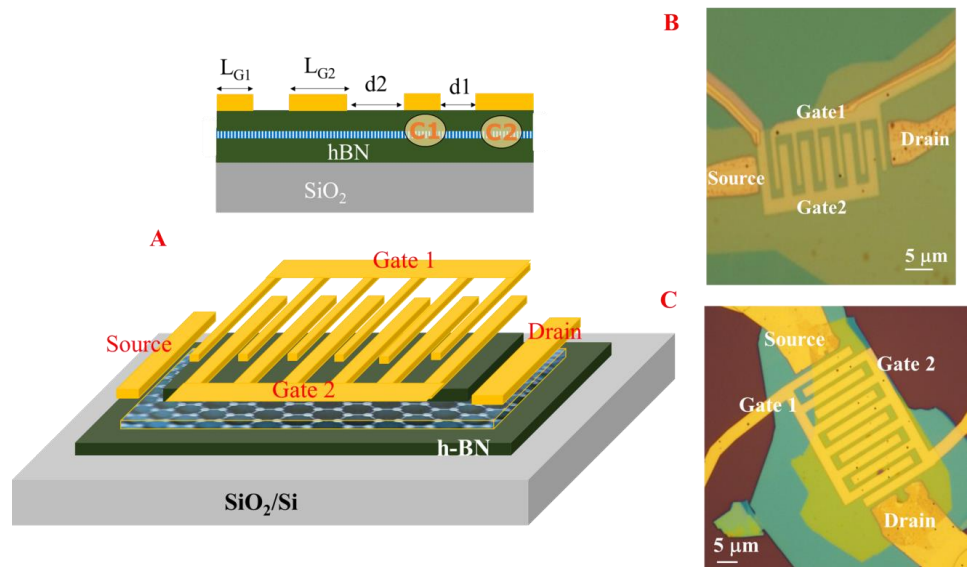

*Fig. 1. Schematic view and device images of the active graphene metamaterials. Schematic illustrating of the hBN/graphene/hBN heterostructure including the asymmetric dual-grating-gate metallization (A). Optical image of the asymmetric dual-grating-gate A- DGG1 (B) and the symmetric S-DGG (C) samples.*

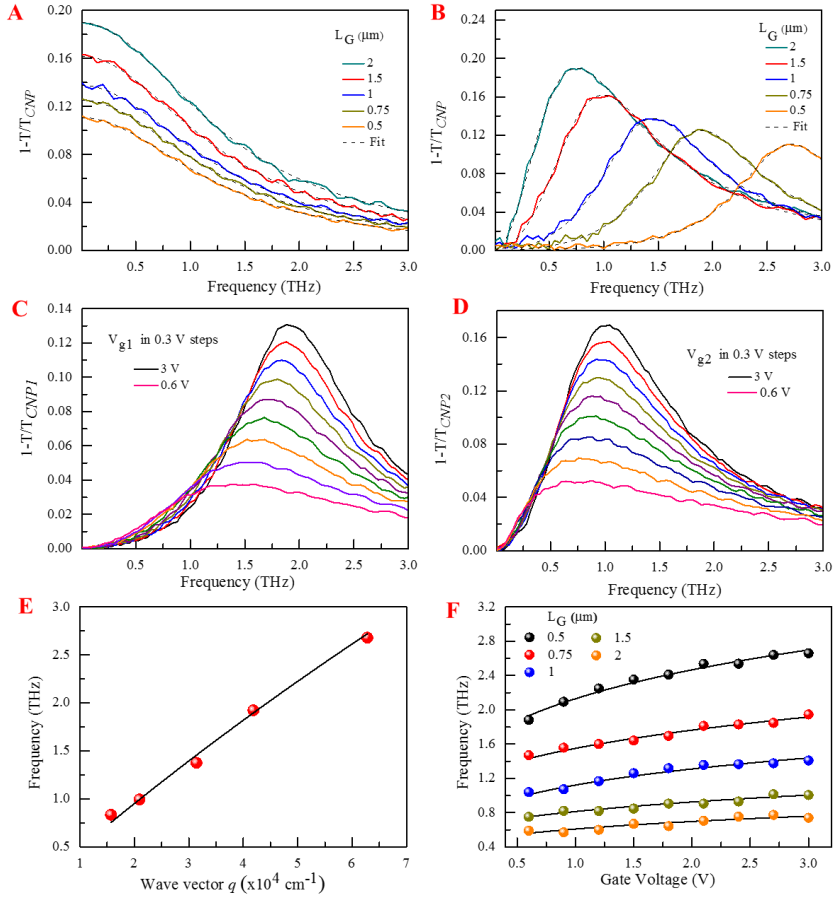

*Fig. 2. Gate voltage and length dependent extinction spectra of the graphene structures for $V_d = 0$ V.* Plasmonic cavity length ($L_G$) dependent extinction spectra of the three devices with incident light polarized parallel (**A**) and perpendicular (**B**) to the gates fingers. The measured line shape is well reproduced by a Drude model fit for parallel polarization (dashed line in (**A**)) and a damped oscillator model fit for perpendicular polarization (dashed line in (**B**)). Measured gate voltage-dependent transmission spectra $1-T/T_{CNP}$ of the asymmetric device A-DGG1 with incident light polarized perpendicular to the gates fingers, with biased cavities C1 when sweeping $V_{g1}$ while the voltage on the other gate electrode is kept constant at $V_{g2} = V_{CNP2}$ (**C**) and biased cavities C2 when sweeping $V_{g2}$ while the voltage on the other gate electrode is kept constant at $V_{g1} = V_{CNP1}$ (**D**). Scaling laws of graphene plasmon resonance frequency in the three devices as a function of wave vector $q = \pi/L_G$ (**E**) and gate voltage (**F**) and fits to data are shown as solid lines in **E** and **F**.

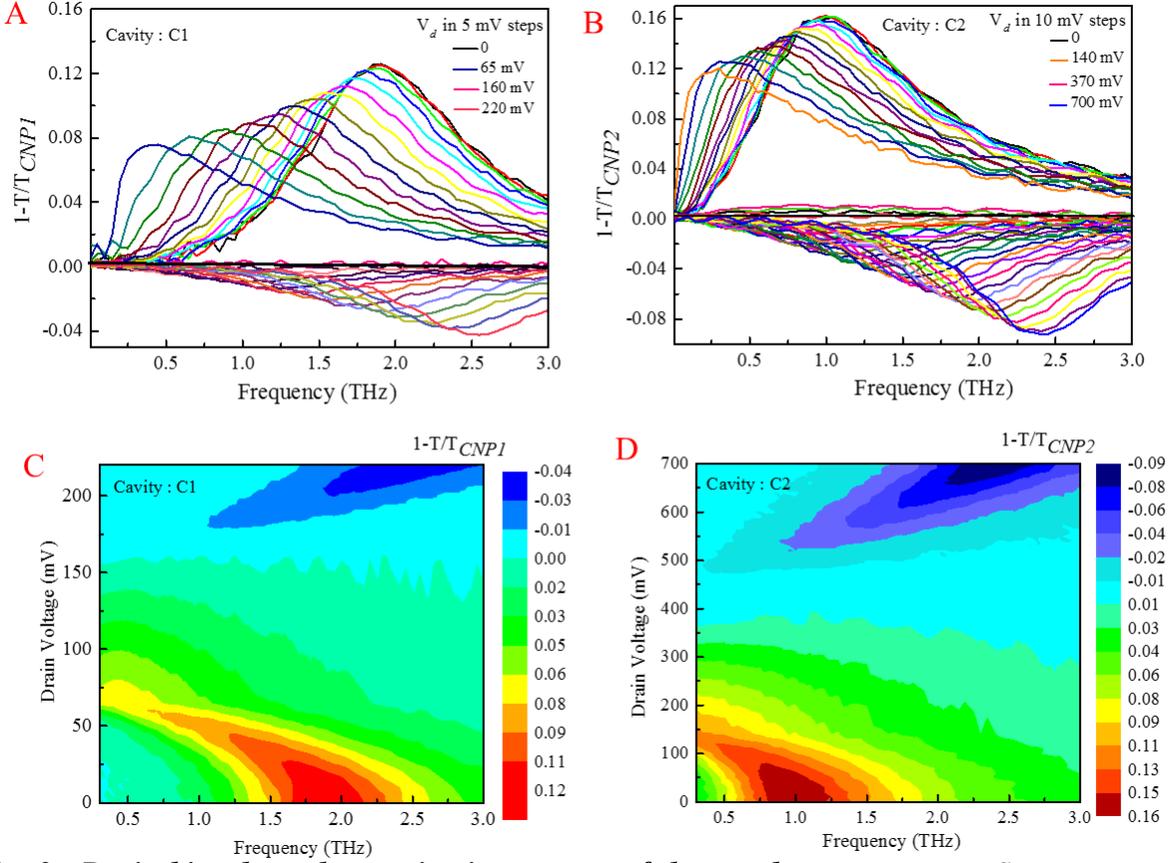

*Fig. 3. Drain bias dependent extinction spectra of the graphene structures.* Spectra measured in cavities C1 of device A-DGG1 for fixed $V_{g1} - V_{CNP1} = 3$ V and $V_{g2} = V_{CNP2}$ (**A**), and cavities C2 of the same device for fixed $V_{g2} - V_{CNP2} = 3$ V and $V_{g1} = V_{CNP1}$ (**B**) when varying $V_d$. Contour plot of the experimental extinction spectra as a function of frequency and $V_d$ measured in A-DGG1 with the biasing conditions mentioned above for cavities C1 (**C**) and C2 (**D**).

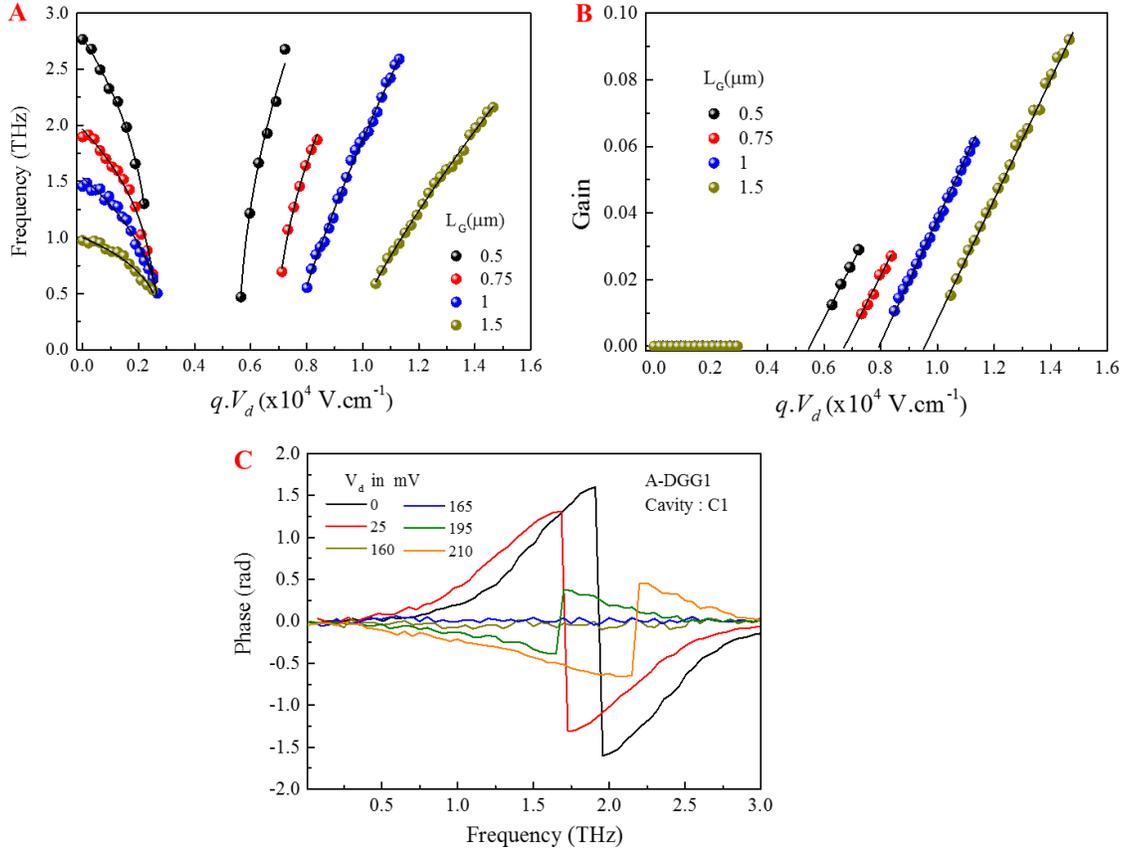

*Fig. 4. **Drain bias dependent resonance frequencies, gain and phase spectra.** Scaling laws of the measured graphene plasmon resonance frequency in the three devices as a function of $qV_d$ and fits to data shown as solid lines (**A**). Maximal gain coefficient (minima of the negative extinction spectra) measured in devices A-DGG1 and A-DGG2 (**B**). Clear threshold like behavior is seen with onset field intensities of ∼ 5.4 kV/cm (black dotes), ∼6.6 kV/cm (red dotes), ∼ 7.8 kV/cm (blue dotes) and ∼ 9.4 kV/cm (dark yellow dotes) in the four cavities. Phase change retrieved from the THz electric field in the case of A-DGG1 cavities C1 (**C**). In the region where the negative extinction is observed, the phase shows an inverted behavior compared to that observed in the absorption region.*

**Supplementary Materials**

**Section S1: Samples description.**

Here we investigate dc current driven plasmonic instability in high mobility graphene based active metamaterials. Plasmon modes in homogeneous graphene sheets cannot be optically excited by free space electromagnetic radiation because of the momentum mismatch between free photons and the two-dimensional (2D) plasmons. To enable the coupling between the photons and plasmons one needs to break translation invariance by engineering graphene structures with characteristic dimensions smaller than the light wavelengths. One alternative is to use a periodic grating gate structure positioned above the graphene sheets as used elsewhere in 2D semiconductors (22,35–42). The grating gate modulates the incoming electromagnetic wave and defines the plasmonic vector.

In order to ensure high carriers mobility in our devices, a hBN/monolayer-graphene/hBN heterostructure was deposited onto a SiO$_2$/Si wafer and the top gates were fabricated with a ∼ 40-nm thick hBN gate oxide (thickness estimation from optical color map). Three devices structures have been designed: i) two structures include asymmetric metallization scheme in dual-grating-gate (DGG) alignment with gate fingers width of L$_{G1}$ = 0.75 µm and L$_{G2}$ = 1.5 µm (L$_{G1}$ = 0.5 µm and L$_{G2}$ = 1 µm) separated by d1 = 0.5 µm and d2 = 1 µm gaps (d1 = 0.5 µm and d2 = 2 µm) referred as A- DGG1 shown in Fig. S1B (A- DGG2 shown in Fig. S1C) respectively and ii) one symmetric device structure (S-DGG) with gate fingers width of L$_{G1}$= L$_{G2}$ = 2 µm separated by d1= d2 = 1 µm gaps (shown in Fig. S1A). The gate–voltage V$_{g2}$ –dependent resistance of sample A- DGG1 is shown in Figure S1D. The three devices have electron mobilities around 50.000 cm$^2$/Vs and charge neutral points (CNP) for top gates at V$_{CNP1}$ = V$_{CNP2}$ = − 0.1 V for sample S-DGG; V$_{CNP1}$ = − 0.12 V (V$_{CNP1}$ = 0.15 V) and V$_{CNP1}$ = − 0.06 V (V$_{CNP2}$ = 0.1 V) for sample A-DGG1 (A- DGG2), respectively.

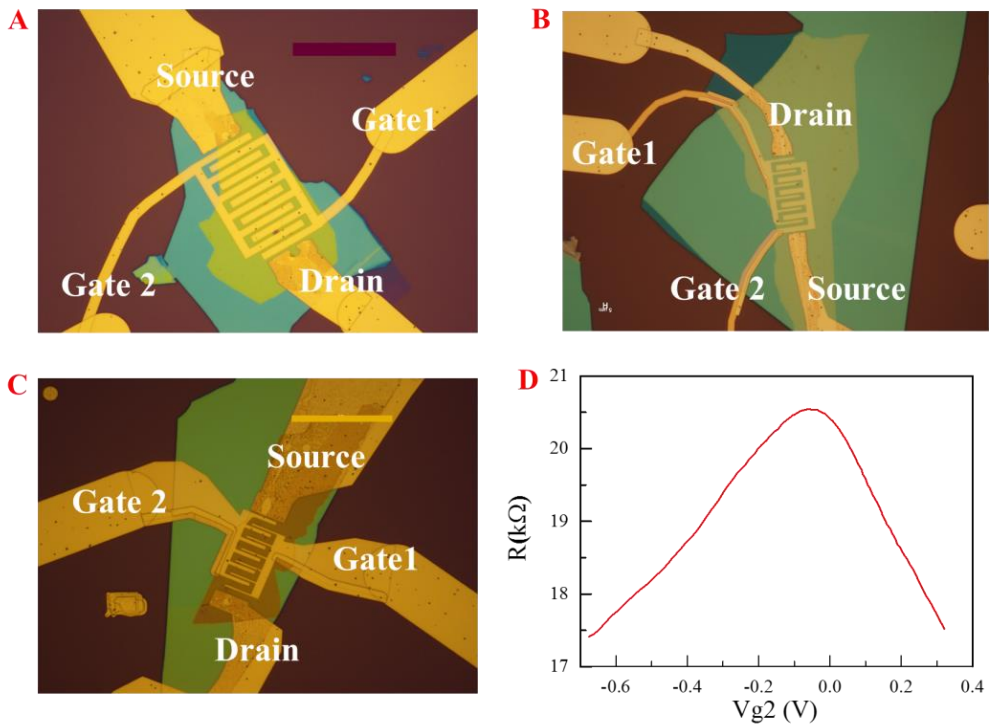

*Figure S1: Device images and the gate dependent resistance of active graphene metamaterials. Optical image of the symmetric S-DGG (A) and asymmetric A-DGG1 (B), A-DGG2 (C) samples. Gate-controlled channel resistance of the sample A-DGG1 (D).*

**Section S2: Supplementary gate voltage results.**

Terahertz time-domain spectroscopy (THz-TDS) was employed to measure the changes in THz pulses transmitted through the graphene plasmonic cavities. The experimental results obtained with the three plasmonic devices were found to be very reproducible. While the gate-voltage dependent data obtained with the device A-DGG1 are shown in the main paper, Fig. S2 depicts the experimentally measured extinction spectra (1-T/T$_{CNP}$) in the asymmetric device A-DGG2 (Figs.S2, A and B) and the symmetric device S-DGG (Fig.S2 C) with plasmonic cavities C1 (while the C2 cavities are in the CNP condition) and C2 (while the C1 cavities are in the CNP condition) for V$_d$ = 0 V. The polarization of the incident THz radiation was set perpendicular to the gates fingers.

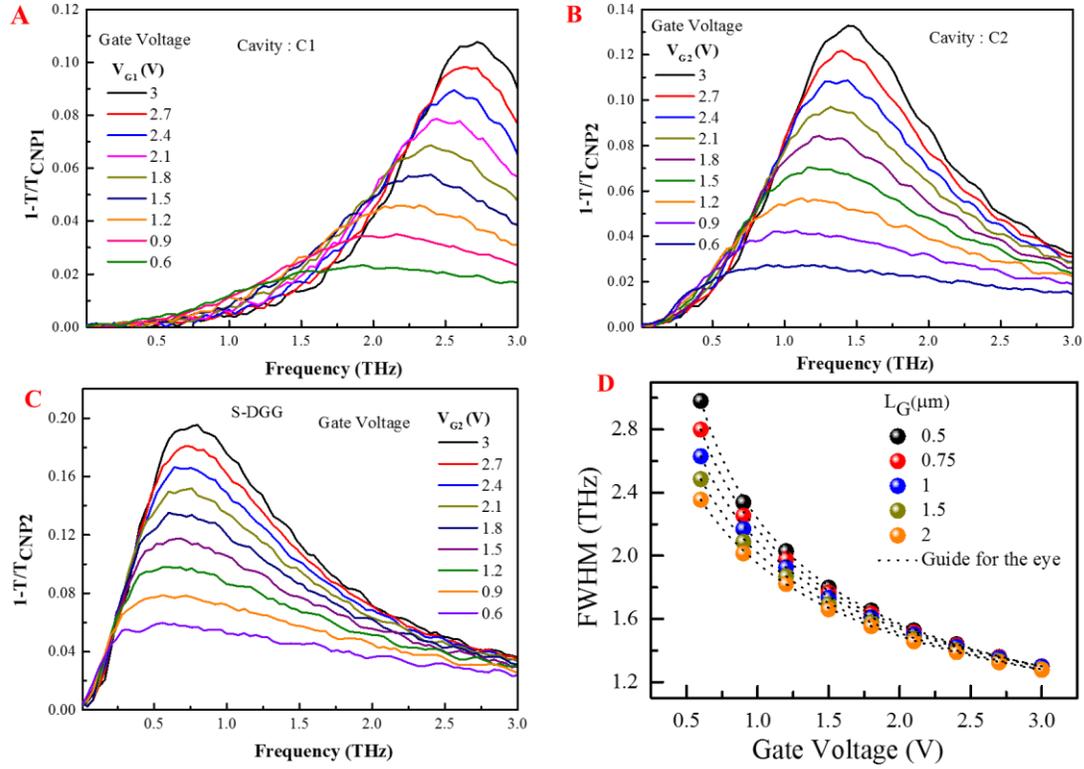

*Figure S2: **Supplementary data, gate voltage dependent extinction spectra of the graphene structures**. Measured gate voltage-dependent transmission spectra 1-T/T$_{CNP}$ of the asymmetric device A-DGG2 (**A** and **B**) and the symmetric device S-DGG (**C**) with biased cavities C1 when sweeping V$_{g1}$ while the voltage on the other gate electrode is kept constant at V$_{g2}$ = V$_{CNP2}$ (**A**) and biased cavities C2 when sweeping V$_{g2}$ while the voltage on the other gate electrode is kept constant at V$_{g1}$ = V$_{CNP1}$ (**B**). Gate-voltage-dependent resonance line-width (FWHM) extracted from the measured extinction spectra for the three samples (**D**). The source to drain voltage V$_d$ =0 V.*

In the quasi-static approximation, the dispersion law for gated 2D plasmons in graphene is determined as (14,23,33):

$$\omega(\omega - \frac{i}{\tau}) = q \frac{4e^2 V_F \sqrt{\pi n}}{\hbar \varepsilon (1 + \coth(qd))} \tag{S1}$$

where $q$ is the plasmon wave vector, $d$ is the gate oxide thickness. The media above and below graphene has the dielectric constant $\varepsilon$, V$_F$ is the graphene Fermi velocity, $n$ is the charge carrier density, $e$ is the elementary charge and $\hbar$ is the reduced Planck's constant. The plasmon resonance frequency is given by (14,33):

$$\omega_P = \left[ q \frac{4e^2 V_F \sqrt{\pi n}}{\hbar \varepsilon (1 + \coth(qd))} \right]^{\frac{1}{2}} \tag{S2}$$

The quantity $qd$ determines the metallic gate screening effects. For strong screening ($qd \ll 1$) the gated plasmon has a linear dispersion law (23) $\omega_P = sq$ where $s = (4\pi^{1/2}n^{1/2}e^2 V_F d / \hbar\varepsilon)^{1/2}$ is the plasmon velocity. In the ungated channel limit ($qd \to \infty$, $\coth(qd) \to 1$) we recover the square root dispersion law of ungated plasmon (14).

Here we assume that the applied gate voltage and the charge density in the channel are related by the Poisson equation, which in the gradual channel approximation is given by $e(n - n_0) = C(V_g - V_{CNP})$, where $n_0$ is the residual carrier concentration induced by density fluctuation due to charged impurities near the Dirac point, $C = \varepsilon/4\pi d$ is an effective electrostatic capacitance per unit area. In the strong screening limit and linear dispersion law we estimate the plasmon velocity to $s \sim 3\times 10^8$ cm/s.

We used a damped oscillator ($\mathrm{Im}(-\omega/(\omega^2 - \omega_P^2 + i\omega\Gamma_P))$) and Drude ($\mathrm{Im}(-1/(\omega + i\Gamma))$) models to describe the line-shape of the extinction spectra for perpendicular and parallel polarized incident THz radiation (dashed lines in Figs.2, A and B of the main paper), where $\Gamma_P$ is the line-width of the plasmon resonance and $\Gamma$ the Drude scattering rate (11,12,30,48). The equation S2 was used to fit the resonance peak extracted from the measured extinction spectra (solid line in Fig.2E of the main paper) with $n_0$ and $d$ as fitting parameters for $V_g - V_{CNP} = \Delta V_g = 3$V, Fermi velocity $V_F = 10^8$ cm/s and $\varepsilon = 6.5$. Good agreement was found for $d = 51.3$ nm which is close to our estimated value of 40 nm of the gate oxide thickness using optical color map, and $n_0 = 4.22\times 10^{10}$ cm$^{-2}$ comparable to the reported values of disorder-induced carrier density fluctuation in high-quality graphene on hBN structures. It is worth mentioning that a quantum capacitance could be considered in series with the dielectric capacitance ($\varepsilon/4\pi d$), in this case the value of $d$ extracted from fits would be even closer to the one estimation from optical color map. To quantify the gate tuning, the resonance frequencies extracted from the extinction spectra (using damped oscillator model) are plotted as a function of the gate voltage for the three samples (Fig.2F of the main paper) and fitted using Eq. S2 (solid lines in Fig.2F of the main paper). Good agreement was found with fitting parameters $n_0 = 4.17$ to $4.51 \times 10^{10}$ cm$^{-2}$ that varies by only $\sim 7.5\%$ and $d = 46.9$ to $52.1$ nm that varies by only $\sim 10\%$ testifying of the goodness of the fits. The $\omega_P \propto n^{1/4}$ power-law dependence confirms that our observed absorption originates from massless Dirac carriers in graphene. Figure S2D depicts gate-voltage-dependent resonance line-width (FWHM) extracted from the fits to the measured extinction spectra for the three samples. Our data show slightly different values of $\Gamma_P$ to that of the Drude scattering rate, most likely due to the contribution in the plasmon broadening of other processes such as the radiative plasmon decay which depends on the engineered coupling radiative antenna structure.

**Section S3: Supplementary drain voltage dependent results.**

The drain voltage dependent experimental results obtained with the three plasmonic devices were also found to be very reproducible. While the data obtained with the device A-DGG1 are shown in the main paper, Fig. S3 depicts the drain-voltage-dependent extinction spectra measured in cavities of the asymmetric device A-DGG2 (Figs. S3, A, B, C and D) and the symmetric device S-DGG (Fig. S3E) with the electrical doping in the cavities C1 at $V_{g1} - V_{CNP1} = 3$ V and $V_{g2} = V_{CNP2}$ shown as $1-T/T_{CNP1}$ (Figs. S3, A and B) and electrical doping in the cavities C2 at $V_{g2} - V_{CNP2} = 3$ V and $V_{g1} = V_{CNP1}$ shown as $1-T/T_{CNP2}$ (Figs. S3 C and D). The THz radiation polarization was set perpendicular to the gates fingers. For the symmetric device S-DGG the electrical doping is set at $V_{g1} - V_{CNP1} = 3$ V and $V_{g2} = V_{CNP2}$ (E). Like in the asymmetric devices we observed a plasmonic resonance absorption peak shifting toward low frequencies as the drain bias increases. However, unlike the asymmetric devices, no amplification was observed in the symmetric devices within the experimental conditions. This most likely highlight the importance of having asymmetric cavity to efficiently observe emission from plasmonic instabilities in our devices (17).

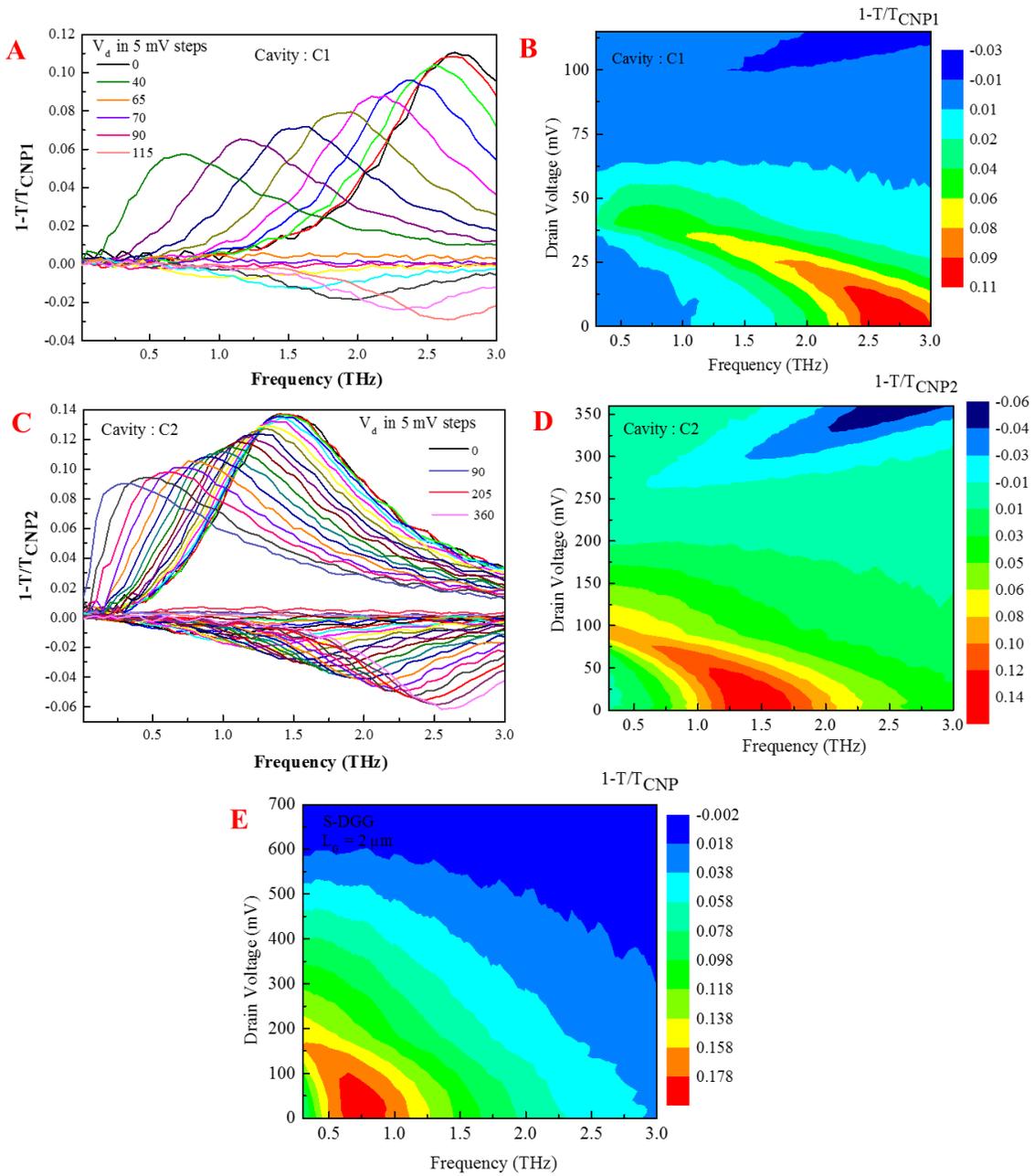

*Figure S3: Supplementary data, drain bias dependent extinction spectra of the graphene structures.* Spectra measured in cavities C1 of device A-DGG2 for fixed $V_{g1} - V_{CNP1} = 3$ V and $V_{g2} = V_{CNP2}$ (*A*), and cavities C2 of the same device for fixed $V_{g2} - V_{CNP2} = 3$ V and $V_{g1} = V_{CNP1}$ (*C*) when varying $V_d$. Contour plot of the experimental extinction spectra as a function of frequency and Vd measured in A-DGG2 (*B* and *D*) and the symmetric sample S-DGG (*E*) with the biasing conditions mentioned above for cavities C1 and C2.